\documentclass[twocolumn,showpacs,superscriptaddress,aps,prl]{revtex4-2}
\usepackage{bm}
\usepackage{amssymb,amsmath}
\usepackage{graphicx}
\usepackage{color}
\usepackage[colorlinks=true,urlcolor=blue,citecolor=blue]{hyperref}

\begin{document}
\title{Interwoven long-range order induced by random fields}
\author{Jeremiah Bender}
\affiliation{Department of Physics, Missouri University of Science and Technology, Rolla, MO 65409, USA}
\author{Thomas Vojta}\email[Contact author: ]{vojta@mst.edu}
\affiliation{Department of Physics, Missouri University of Science and Technology, Rolla, MO 65409, USA}

\begin{abstract}
We propose a distinct type of long-range ordered phase that can occur in classical and quantum many-particle systems. It is induced by impurities and defects that locally break a subset of the order-parameter symmetries, i.e., by random-field disorder that couples to a composite vestigial order parameter. The proposed ``implectic'' phase is characterized by spontaneous symmetry breaking on the background of the spatially interwoven domain structure created by the random fields. We explicitly demonstrate the existence of this phase in a layered $J_1-J_2$ Ising magnet by means of large-scale Monte Carlo simulations. We then discuss numerous potential applications in systems featuring charge and spin density wave order including frustrated magnets, cuprate and iron-based superconductors, and ultracold atoms.
\end{abstract}

\date{\today}

\maketitle


\paragraph{Introduction:} The search for new phases of matter is one of the central themes of physics. Traditionally, different phases or long-range ordered states have been understood in terms of their symmetries. Landau's paradigm of spontaneous  symmetry breaking \cite{Landau37b,Landau37d,LandauLifshitz_book80} provides a systematic framework of classifying such ordered states. However, not all ordered phases follow Landau's paradigm; topologically ordered states are not characterized by symmetry breaking but instead feature long-range quantum entanglement and fractional excitations \cite{Wen13,Wen17,Sachdev19}.

Correlated quantum materials often have complex phase diagrams that exhibit multiple intertwined ordered states \cite{FradkinKivelsonTranquada15}. These can arise, e.g., when the system features a primary ordered phase that beaks several symmetries as well as one or more vestigial phases that break only subsets of these symmetries \cite{FernandesOrthSchmalian19}. Examples of this mechanism include the vestigial Ising degree of freedom  found in a frustrated Heisenberg magnet \cite{ChandraColemanLarkin90}, the spin nematic phase \cite{AndreevGrishchuck84}, and the electronic nematic phase in iron-based superconductors \cite{Fangetal08,XuMullerSachdev08,Fernandesetal12}.

Realistic materials usually contain random vacancies, impurities, and other kinds of quenched disorder. Disorder can also be introduced artificially into intrinsically clean systems such as ultracold atomic gases. Quenched disorder can dramatically influence phases and phase transitions of many-particle systems, see, e.g. Refs.\ \cite{Vojta06, Vojta13, Vojta19} for recent reviews. Particularly strong effects arise when the disorder \emph{locally} breaks the symmetry between equivalent macroscopic states while preserving the symmetry globally (in the statistical sense). As this type of disorder corresponds to a random external field in a magnetic system, it is usually called random-field disorder. A striking example of a random-field magnet is realized in {LiHo$_x$Y$_{1-x}$F$_4$} \cite{TGKSF06,SBBGAR07,Schechter08}. Random-field disorder is ubiquitous in nature. For  example, it generically occurs when the ordered phase breaks a \emph{real-space} symmetry \cite{Vojta19} such as in nematic liquid crystals in porous media \cite{MCBB94} and in stripe states of high-temperature superconductors \cite{CDFK06}.

Building on heuristic arguments by Imry and Ma \cite{ImryMa75}, Aizenman and Wehr \cite{AizenmanWehr89} showed rigorously that (arbitrarily weak) random-field disorder prevents spontaneous symmetry breaking in dimensions $d\le2$ for discrete order parameter symmetry and $d\le4$ for continuous symmetry because the system forms finite-size domains that align with the local random field. In higher dimensions, the uniformly ordered state survives for
weak random fields,  but domains form for sufficiently strong random fields. A quantum version of the Aizenman-Wehr theorem was proven by Greenblatt et al.
\cite{GreenblattAizenmanLebowitz09,AizenmanGreenblattLebowitz12}.

In this Letter, we show that a distinct long-ranged ordered phase can be induced, in three or more space dimensions, by random-field type disorder that locally breaks only a subset of the symmetries broken by the primary order parameter. This means the random field couples to a composite vestigial order parameter rather than the primary order parameter.
According to Imry and Ma \cite{ImryMa75}, these random fields (if sufficiently strong) lead to the formation of finite-size domains of the vestigial composite order.   Conventional wisdom suggests that such domains prevent spontaneous  symmetry breaking and thus destroy long-range order.
This is indeed the case in (up to) two space dimensions.  However, in three (and higher) dimensions, the domains can form interwoven, percolating random networks in space. Each of these networks can  undergo a sharp phase transition that breaks the remaining order parameter symmetries that are not already broken by the random fields. The resulting long-range ordered phase, which we call an ``implectic'' phase  (\emph{implexus} is Latin for interwoven), is  characterized by the coexistence of several ordered states on these interwoven random networks.

The remainder of this Letter is organized as follows. As a prototypical example, we first analyze a layered frustrated $J_1-J_2$ Ising magnet with a nematic random field. Employing large-scale Monte Carlo simulations, we verify the existence of the proposed interwoven, implectic long-range ordered phase. We also identify the sharp phase transition separating it from the paramagnetic high-temperature phase.  We then consider the conditions under which this state can be observed in realistic systems, and we discuss several potential applications  in quantum materials and beyond.

\begin{figure}
\includegraphics[width=\columnwidth]{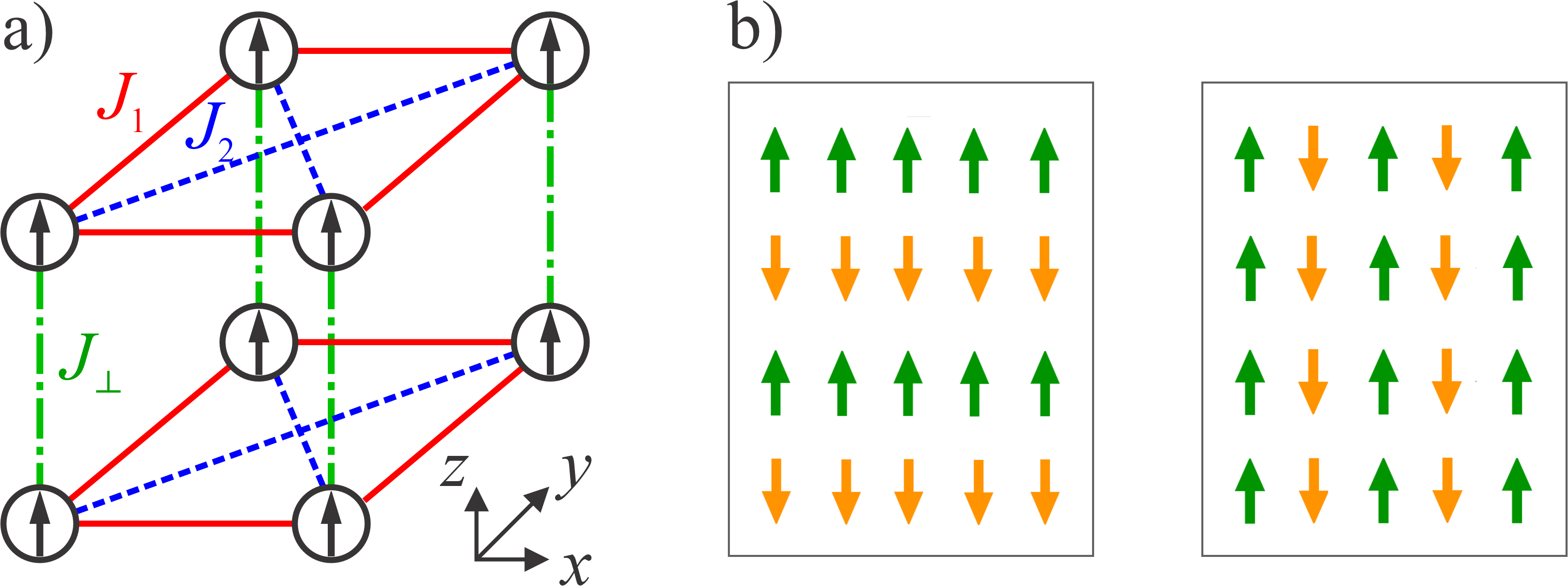}
\caption{a) Interactions of the layered $J_1-J_2$ Ising model. b) Schematic of the stripe order in the $xy$ plane in the low-temperature phase for $|J_2|/J_1>1/2$ in the absence of quenched disorder. Stripes can be aligned with either the $x$  direction or the $y$ direction.}
\label{fig:lattice}
\end{figure}

\paragraph{$J_1-J_2$ Ising magnet with nematic random field:} We start with the layered frustrated $J_1-J_2$ Ising model, a three-dimensional generalization of the well-known square-lattice $J_1-J_2$ Ising model. It is defined on a cubic lattice of $N=L^3$ sites and given by the Hamiltonian
\begin{equation}
H_0 = - J_1 \sum_{\langle ij \rangle} S_i S_j  - J_2 \sum_{\langle\langle ij \rangle\rangle}  S_i S_j - J_\perp \sum_{\langle ij \rangle_\perp}  S_i S_j~.
\label{eq:H0}
\end{equation}
Here, $S_i = \pm 1$ is a classical Ising spin, $\langle ij \rangle$ denotes pairs of nearest-neighbor
sites in the $xy$-plane,  coupled by the ferromagnetic interaction $J_1>0$; $\langle\langle ij \rangle\rangle$
denotes next-nearest neighbor pairs  in the $xy$-plane, coupled by the antiferromagnetic interaction $J_2 < 0$,
and $J_\perp >0$ is the inter-layer coupling in the $z$-direction (see Fig.\  \ref{fig:lattice}).
The phases of this system are well-understood (see, e.g., Refs.\ \cite{JinSenSandvik12,JinSenGuoSandvik13,KalzHoneckerMoliner11,KalzHonecker12,GodoySchmidtZimmer20,Sorokin18,Sorokin22} and references therein).
It displays paramagnetic behavior at high temperatures.  As the temperature is lowered, two distinct
long-range ordered phases appear.
For $|J_2|/J_1 < 1/2$, the low-temperature phase is ferromagnetic. We are interested in the case $|J_2|/J_1>1/2$  for which the low-temperature
phase features a striped spin order in the $xy$ plane that is repeated in the $z$ direction. Stripes can be aligned with either the $x$  direction or the $y$ direction;
thus this phase breaks not only the $Z_2$ Ising spin symmetry but also the $C_4$ rotation symmetry of the
$xy$ square lattice. The stripe order can be characterized by a (primary) two-component order parameter $\psi=(\psi_x,\psi_y)$ with
\begin{equation}
\psi_x = \frac 1 N \sum_i (-1)^{y_i} S_i ~, \quad\psi_y = \frac 1 N \sum_i (-1)^{x_i} S_i ~.
\label{eq:stripe_OP}
\end{equation}
Here, $x_i$ and $y_i$ are the (integer) coordinates of site $i$. The (composite) vestigial nematic order parameter \cite{FernandesOrthSchmalian19}
\begin{equation}
\eta =\frac 1 N \sum_i \eta_i = \frac 1 {4N} \sum_i S_i( S_{i,+x} + S_{i,-x} - S_{i,+y} -  S_{i,-y}  )
\label{eq:nematic_OP}
\end{equation}
characterizes the breaking of the $C_4$ lattice symmetry, irrespective of the presence or absence of magnetic long-range order  \footnote{See Supplemental Material at XXX with Refs.\ \cite{MHSS96,Wolff89,SwendsenWang87,MRRT53,HukushimaNemoto96,EarlDeem05} for details of the order parameter structure and the Monte Carlo simulations.}. Here, $S_{i,+x}$ refers to the nearest neighbor of spin  $S_i$ in the positive $x$ direction.

We now introduce quenched disorder that locally favors one stripe direction over the other. Such disorder can arise, e.g., via site dilution as vacancy pairs on nearest neighbor sites break the symmetry between the two stripe directions \cite{KunwarSenVojtaNarayanan18,YeNarayananVojta22}. Alternatively, random strains in the sample lead to local differences between the interactions in the $x$ and $y$ directions \cite{MeeseVojtaFernandes22,Lietal24}. Importantly, such disorder locally breaks the symmetry between the stripe directions but \emph{not} the Ising spin symmetry. It can thus be modeled as a quenched random field $\phi_i$ that couples to the local nematic order parameter rather than the primary stripe order parameter, yielding the Hamiltonian
\begin{equation}
H= H_0 +\sum_i \phi_i \eta_i~.
\label{eq:H}
\end{equation}


\paragraph{Simulations:} We analyze the effects of the nematic random field on the phases of the layered $J_1-J_2$ Ising model by means of large-scale Monte Carlo simulations. For definiteness, we fix the interactions at  $J_1=-J_2=J_\perp=1$. The $\phi_i$ are uniformly distributed on the interval $(-W,W)$. They are short-range correlated with a correlation length $\xi_d$ \cite{Note1}. Monte Carlo simulations of random-field systems are hampered by slow equilibration. To overcome this difficulty, we implement an efficient parallel tempering algorithm \cite{SwendsenWang86,Geyer91,MarinariParisi92}, for details see the Supplemental Material \cite{Note1}.

\begin{figure}
\includegraphics[width=0.95\columnwidth]{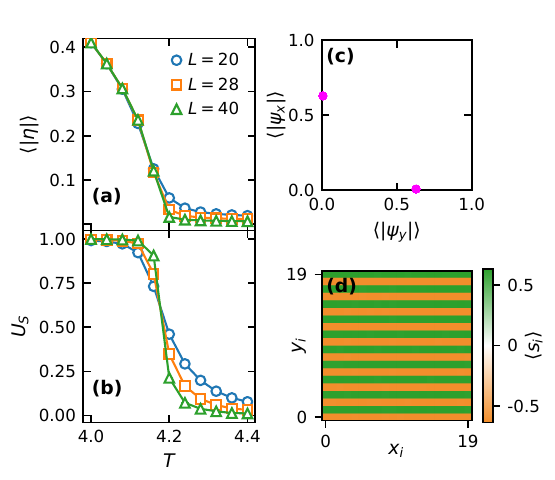}
\caption{Simulation results for the layered $J_1-J_2$ Ising model without random field. (a) Nematic order parameter $\langle |\eta|\rangle$ and (b) stripe Binder cumulant $U_s$  vs.\ temperature $T$ from different system sizes $L$. (c) $\langle| \psi_x |\rangle  $ vs.\ $ \langle | \psi_y | \rangle$ for 100 individual samples at $T=4.0$. (d) Local magnetization $\langle S_i \rangle$ in a single layer of one sample at $T=4.0$. Statistical errors in panels a) and (b) are below the symbol size.}
\label{fig:clean}
\end{figure}

For reference and to test our numerical algorithms, we first performed simulations for zero random field, $W=0$. Figure \ref{fig:clean} presents a summary of the results.
In agreement with the literature \cite{Sorokin18}, the system features a phase transition at $T_c \approx 4.18$ from a paramagnet at high temperatures to a stripe state at low temperatures. This is indicated by the onset of a nonzero nematic order parameter $\langle \eta \rangle$ as well as the crossing of the stripe Binder cumulant $U_s= 1 - \langle \psi^4\rangle/ (2\langle \psi^2\rangle^2)$. Here $\langle \ldots \rangle$ denotes the thermodynamic average. The stripe order is uniform across the entire sample as can be seen in Fig.\  \ref{fig:clean}(c); it shows, as dots, the values of the stripe order parameter components $\langle \psi_x \rangle$ and $\langle \psi_y \rangle$ for several independent samples at the lowest temperature. Each sample either has $\langle \psi_x \rangle =0$ and $\langle \psi_y \rangle \ne 0$ or vice versa, indicating a spontaneous breaking of the lattice symmetry. The local magnetization in a single layer of one sample is shown in panel (d).

\begin{figure}
\includegraphics[width=\columnwidth]{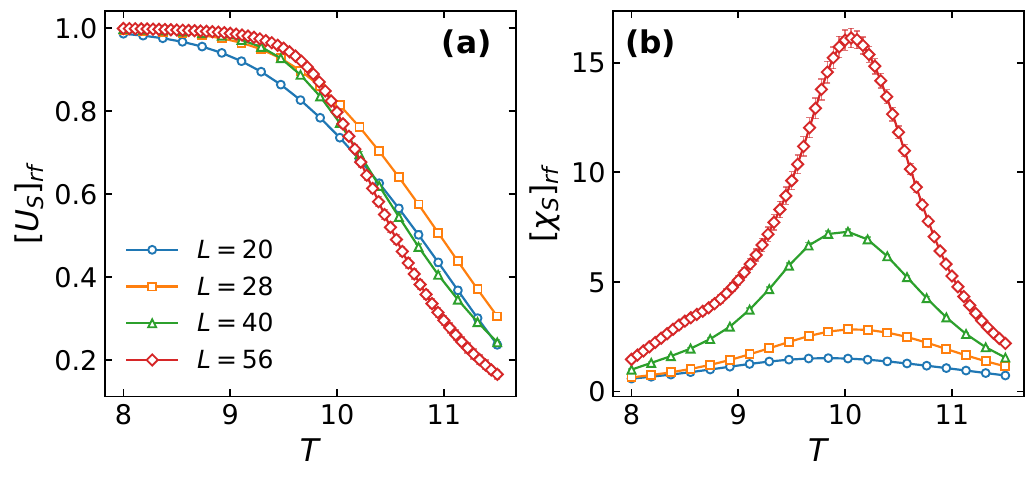}
\caption{(a)  Stripe Binder cumulant $[U_s]_\mathrm{rf}$ and (b) stripe susceptibility $[\chi_s]_\mathrm{rf}$ vs.\ temperature $T$. Random field strength $W=8$ and correlation length $\xi_d=3$. The data are averaged over 1000 random-field configurations for system sizes $L=20, 28, 40$ and 256 configurations for $L=56$. Statistical errors are shown but mostly smaller than the symbol size.}
\label{fig:Us_chi}
\end{figure}

\begin{figure}
\includegraphics[width=\columnwidth]{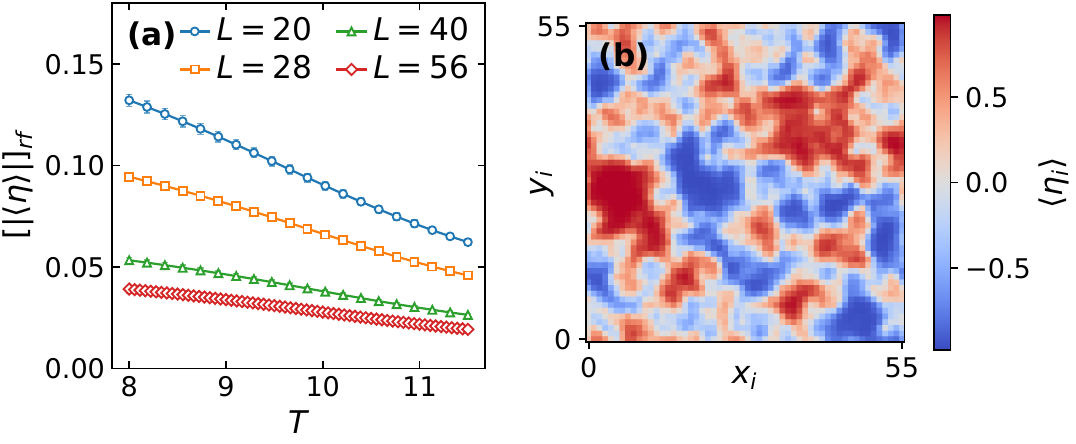}
\caption{(a) Nematic order parameter $|\langle \eta \rangle|$ vs.\ temperature $T$ at $W=8$ and $\xi_d=3$, averaged over 1000 random-field configurations for system sizes $L=20, 28, 40$ and 256 configurations for $L=56$. (b) Local nematic order parameter $\langle \eta_i \rangle$ in a single layer of one sample at $T=8.0$. For $\langle \eta_i \rangle >0$, stripes locally align with the $x$ direction, while $\langle \eta_i \rangle <0$ corresponds to stripes aligned with the $y$ direction.}
\label{fig:eta}
\end{figure}

We now turn to the behavior of the system in the presence of a strong nematic random field. Figure \ref{fig:Us_chi} shows the stripe Binder cumulant $[U_s]_\mathrm{rf}$ and the stripe susceptibility $[\chi_s]_\mathrm{rf} = [(\langle \psi^2 \rangle - \langle |\psi| \rangle^2)/T]_\mathrm{rf}$ as functions of temperature for random-field strength $W=8$ and correlation length $\xi_d=3$. (Here, $[\ldots ]_\mathrm{rf}$ denotes the average over random-field configurations.)
The data clearly show a transition into a long-range stripe-ordered state at $T_c \approx 10$, as indicated by the crossing of the Binder cumulants and the divergence of the susceptibility with system size $L$. On the other hand, the nematic order parameter $[\langle \eta \rangle]_\mathrm{rf}$ is small in the same temperature range and decays rapidly with $L$, as can be seen in Fig.\ \ref{fig:eta}(a).
This indicates the absence of long-range nematic order. Figure \ref{fig:eta}(b) shows the local nematic order parameter $\langle \eta_i \rangle$ in a single layer of one sample, illustrating the destruction of nematic order by the formation of domains of a characteristic size of about 10 lattice constants.

At first glance, the results in Figs.\ \ref{fig:Us_chi} and \ref{fig:eta} seem to contradict each other. In order to understand the unusual low-temperature state,
we need to explain how long-range stripe order can emerge in the presence of finite-size nematic domains. The key insight is that the domains favoring $x$ stripes and the domains favoring $y$ stripes form two separate but interwoven random networks in space. In three dimensions, both of these networks can percolate and span the entire lattice. This implies that the spins  on each of the networks can undergo a true phase transition that spontaneously breaks the remaining Ising symmetry and establishes the implectic phase, i.e., long-range interwoven stripe order ($x$ stripes on one of the networks and $y$ stripes on the other).

\begin{figure}
\includegraphics[width=\columnwidth]{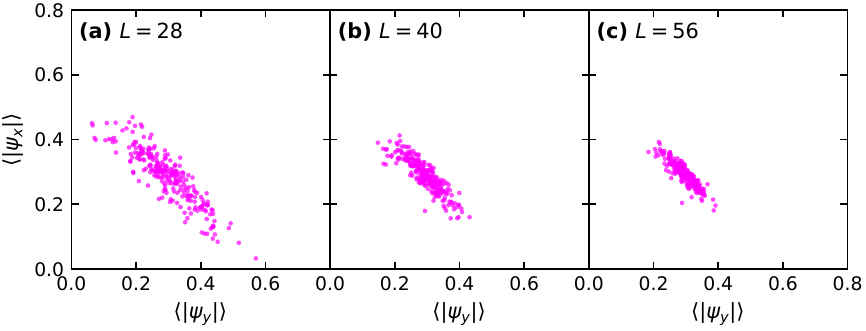}
\caption{$\langle | \psi_x | \rangle$ vs.\ $\langle | \psi_y | \rangle$ for 256 individual samples at $T=8.0$,  $W=8$, and $\xi_d=3$. Each dot represents one individual disorder configuration.}
\label{fig:scatter}
\end{figure}

\begin{figure}
\includegraphics[width=\columnwidth]{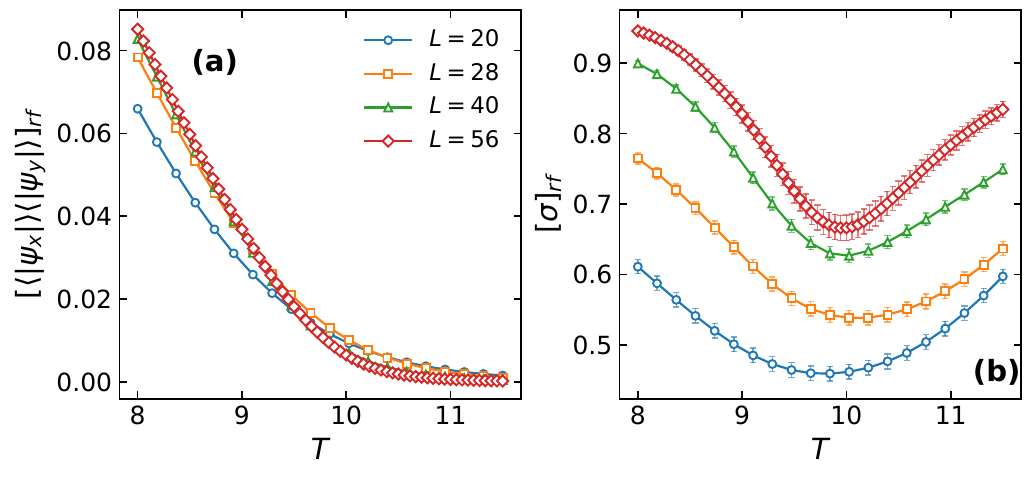}
\caption{(a) Order parameter product $[\langle | \psi_x | \rangle \langle | \psi_y | \rangle ]_\mathrm{rf}$ vs.\ temperature $T$ for  $W=8$, $\xi_d=3$, averaged over 1000 random-field configurations for system sizes $L=20, 28, 40$ and 118 configurations for $L=56$.  (b) Coexistence parameter $[\sigma]_\mathrm{rf} =4 [\langle\psi_x\rangle^2 \langle \psi_y \rangle^2/\langle \psi^2 \rangle^2]_\mathrm{rf} = [1-\cos(4\Theta_\mathrm{coex})]_\mathrm{rf}/2$  vs.\ temperature $T$ for the same parameters.    }
\label{fig:coex}
\end{figure}

In order to verify this intriguing scenario in our system, Fig.\ \ref{fig:scatter} shows the  stripe order parameter components $\langle \psi_x \rangle$ and $\langle \psi_y \rangle$ for several individual random field configurations at the lowest temperature.
In contrast to the clean case ($W=0$), shown in Fig.\ \ref{fig:clean}(c), each sample features nonzero values for both $\langle \psi_x \rangle$ and $\langle \psi_y \rangle$. Importantly, with increasing system size $L$, the distribution of values narrows around the limit $\langle \psi_x \rangle = \langle \psi_y \rangle$, indicating coexistence of $x$ stripes and $y$ stripes on two random networks of equal weight in the thermodynamic limit. To quantify the coexistence
of $x$ stripes and $y$ stripes, we present in Fig.\ \ref{fig:coex}(a) the temperature dependence of $[\langle |\psi_x |\rangle \langle |\psi_y |\rangle ]_\mathrm{rf}$.
This quantity is zero both in the paramagnetic phase and in a uniform stripe state but nonzero if both stripe directions feature coexisting long-range orders. The data show that a
phase with coexisting $x$ and $y$ stripes indeed emerges below $T_c \approx 10$. Fig.\ \ref{fig:coex}(b) shows the coexistence parameter $[\sigma]=  [1-\cos(4\Theta_\mathrm{coex})]_\mathrm{rf}/2$ which parameterizes the angle $\Theta_\mathrm{coex}$ spanned by $\langle \psi_x \rangle$ and $\langle \psi_y \rangle$.
With increasing system size $L$, $[\sigma]_\mathrm{rf}$ approaches its maximum value of unity, corresponding to $\Theta_\mathrm{coex}=\pi/4$, i.e, an equal coexistence of $x$ and $y$ stripes in each sample.

In addition to the simulations with $W=8$ reported above, we have also considered weaker random fields. For $W=1$, the system still shows implectic order, but for $W=0.4$, the low-temperature state instead features uniform stripes order that break the lattice symmetry. This agrees with the prediction of the Imry-Ma criterion: In three dimensions, domain formation is not possible for sufficiently weak random-field disorder.


\paragraph{Discussion:} After having demonstrated the existence of the proposed implectic long-range ordered phase in the layered $J_1-J_2$ Ising model with nematic random field, we now discuss, in more general terms, the conditions under which such states can emerge.

First, the system needs to feature quenched disorder that locally breaks some but not all of the symmetries broken by the primary order parameter of a parent phase, i.e., the disorder needs to act as a vestigial random field that couples to a composite order parameter rather than the primary order parameter. This naturally occurs, e.g., when the primary order parameter breaks a real-space symmetry in addition to a spin, gauge, or other symmetry. Impurities may locally break the real space symmetry but not necessarily the non-real-space symmetry.

Second, the space dimensionality needs to be at least three, such that the random-field induced domains can form several interwoven percolating random networks that all span the entire system. For topological reasons, this is impossible in two dimensions. Correspondingly, the two-dimensional  $J_1-J_2$ Ising model with nematic random field does not show implectic order or any other long-range order \cite{KunwarSenVojtaNarayanan18,YeNarayananVojta22}.

Third, the disorder needs to be sufficiently strong to induce the formation of domains of the composite order parameter. In three dimensions, this means the disorder strength $W$ needs to exceed a threshold $W_c$ that can be estimated from an Imry-Ma argument \cite{ImryMa75,Note1}. This threshold decreases as $1/\xi_d$ with increasing disorder correlation length which implies that random fields with sizable correlations can induce domain formation in three dimensions even if their microscopic energy scale is small.


\paragraph{Applications:} Many quantum materials feature complex ordered phases with multi-component order parameters \cite{FradkinKivelsonTranquada15,FernandesOrthSchmalian19}. Paradigmatic examples include the cuprate high-$T_c$ superconductor materials which often contain significant disorder but nonetheless exhibit sharp phase transitions into long-range ordered phases that intertwine superconductivity, charge-density waves, spin-density waves, and pair-density waves. If these systems have several degenerate ground states that are related by a lattice symmetry (for example, two or more equivalent ordering wave vectors $\mathbf{Q}$), defects may locally break this lattice symmetry, generating a vestigial random field and opening a route to implectic order.

Due to the electron-phonon coupling, random strain in a sample can act as a random field for an electronic nematic order parameter in a variety of quantum materials \cite{FKLEM10}. Random strains decay slowly in space and thus generically create correlated disorder \cite{LandauLifshitz_book70}. Electronic nematicity is often intertwined with a density-wave type orders in cuprates, iron arsenides, heavy fermion compounds, and other quantum materials.  In many of these systems macroscopic experiments observe signatures of a sharp phase transition whereas local probes report nematic domain breakup due to disorder (see, e.g., Ref.\ \cite{MeeseFernandes26} and references therein).

The implectic interwoven order introduced in this Letter may provide an explanation for this seemingly contradictory behavior in at least some of these materials.
For example, GdRhIn$_5$ shows signatures of a magnetic stripe phase, but no orthorhombic distortion is observed in X-ray experiments \cite{Granadoetal06,Betancourthetal19}, and the authors of Ref.\  \cite{Betancourthetal19} suggested that the ordered phase is a spatially inhomogeneous mixture of the two possible degenerate ground states.
 Establishing the implectic scenario beyond doubt in this and other materials requires both (i) the verification of domain formation for the vestigial order parameter and (ii) the identification of a sharp phase transition that breaks the remaining order parameter symmetries and produces long-range order (on each of the interwoven random domain networks)  with a correlation length \emph{larger than the characteristic domain size}.

Intertwined and vestigial phases have also been found in cold atom systems \cite{GopalakrishnanShchadilovaDemler17,Jinetal21,Wangetal23}. Artificially adding disorder to these systems, e.g., via speckles, can provided an alternative route to observe the proposed implectic  ordered state.


\paragraph{Conclusions:} To summarize, we have introduced implectic order, a distinct type of long-range order that is induced by random fields that locally break a subset of the order parameter symmetries of a parent phase and thus couple to a vestigial order parameter. Implectic order is characterized by the coexistence of several long-range ordered states on separate but interpenetrating and percolating random networks formed by the domains of the vestigial order.

We have demonstrated the existence of implectic order for a prototypical model, the layered $J_1 - J_2$ Ising magnet. Without quenched disorder, the low-temperature phase of this model features stripe magnetic order that spontaneously breaks the Ising spin symmetry as well as the symmetry between the $x$ and $y$ directions of the lattice. A nematic random field locally prefers one stripe direction over the other, i.e., it couples to the vestigial nematic order parameter. Our Monte Carlo simulations have shown that (sufficiently strong) random fields prevent nematic long-range order by domain breakup. The domains supporting either $x$ or $y$ stripes form two interpenetrating networks, and the spins on each of these networks undergoes a sharp phase transition that breaks the Ising symmetry, establishing the implectic order.

Table \ref{table:symmetries} shows the symmetries of the various phases of the layered $J_1 - J_2$ Ising magnet with nematic random field.
\begin{table}
\renewcommand*{\arraystretch}{1.2}
	\begin{tabular*}{\columnwidth}{@{\extracolsep{\fill}}lccccc}
		\hline\hline
		 & uniform  & vestigial & implectic & paramagn. \\
             &  stripe     & nematic  &           &                  \\  \hline
		Ising  symmetry & broken    & yes & broken & yes \\
		Lattice symmetry  &broken   & broken & yes & yes\\
         \hline\hline
	\end{tabular*}
	\caption{Phases and their symmetries of the layered $J_1 - J_2$ Ising model with nematic random field. The vestigial nematic phase, while possible on symmetry grounds, is not observed in the $J_1 - J_2$ Ising model.}
	\label{table:symmetries}
\end{table}
The uniform stripe phase, realized at low temperatures  for sufficiently weak random field breaks both the Ising spin symmetry and the lattice symmetry between the $x$ and $y$ directions. A hypothetical vestigial nematic phase would break the lattice symmetry while not breaking the Ising spin symmetry. Such a phase, which could in principle appear between the stripe phase and the paramagnet, is not observed in the  $J_1 - J_2$ Ising model. The implectic phase established in this Letter does break the Ising spin symmetry while the lattice symmetry is not broken globally due to domain formation.

We hope that our results will open up the study of implectic order in many realistic systems. To this end, we have formulated general conditions under which implectic order can appear, and we have identified several potential applications of this concept in quantum materials and beyond.
Our results raise the number of important theoretical and experimental questions. These include: What is the role of spatial anisotropies that are prevalent in many quantum materials? Can one design smoking-gun experiments that distinguish implectic order from simpler scenarios such as macroscopic phase coexistence?  A particularly interesting question is whether implectic order can provide an explanation for the puzzling dichotomy between domains observed by local probes in electronic nematics, and the sharp phase transitions found in many macroscopic experiments.

In this Letter, we have focused on order parameters that break a real-space symmetry in addition to a spin, gauge, or other symmetry. However, we expect implectic order to be possible in a broader class of systems that feature other multi-component order parameters that break more than one symmetry. This includes, for example, certain random-anisotropy magnets. Specifically, the physics of a ferromagnetic three-dimensional four-state clock model with a vestigial random field that locally prefers the clock variable to align with either the $\pm x$ direction or the $\pm y$ direction is analogous to the system discussed above.

\textit{Acknowledgements}---We thank Rafael Fernandes and Jos\'e Hoyos for helpful discussions. The simulations were performed on the Pegasus, Foundry, and Mill clusters at Missouri S\&T.

\textit{Data availability}---The data that support the findings of this article are openly available \cite{Bender26}.


%

\clearpage       

\setcounter{equation}{0}
\setcounter{figure}{0}
\setcounter{table}{0}
\setcounter{page}{1}
\makeatletter
\renewcommand{\theequation}{S\arabic{equation}}
\renewcommand{\theHequation}{S\arabic{equation}}
\renewcommand{\thefigure}{S\arabic{figure}}
\renewcommand{\theHfigure}{S\arabic{figure}}
\renewcommand{\thetable}{S\arabic{table}}
\renewcommand{\theHtable}{S\arabic{table}}
\renewcommand{\bibnumfmt}[1]{[S#1]}
\renewcommand{\citenumfont}[1]{S#1}

\onecolumngrid
\begin{center}
{\large\bf Supplemental Material for: Interwoven long-range order induced by random fields}

\bigskip

Jeremiah Bender $^1$ and Thomas Vojta $^1$

\smallskip

{\it
$^1$Department of Physics, Missouri University of Science and Technology, Rolla, MO 65409, USA\\
}
(Dated: \today)

\vspace*{5mm}
\end{center}
\twocolumngrid


\section{S1. Primary and vestigial order parameters}

\begin{figure}[b]
\includegraphics[width=\columnwidth]{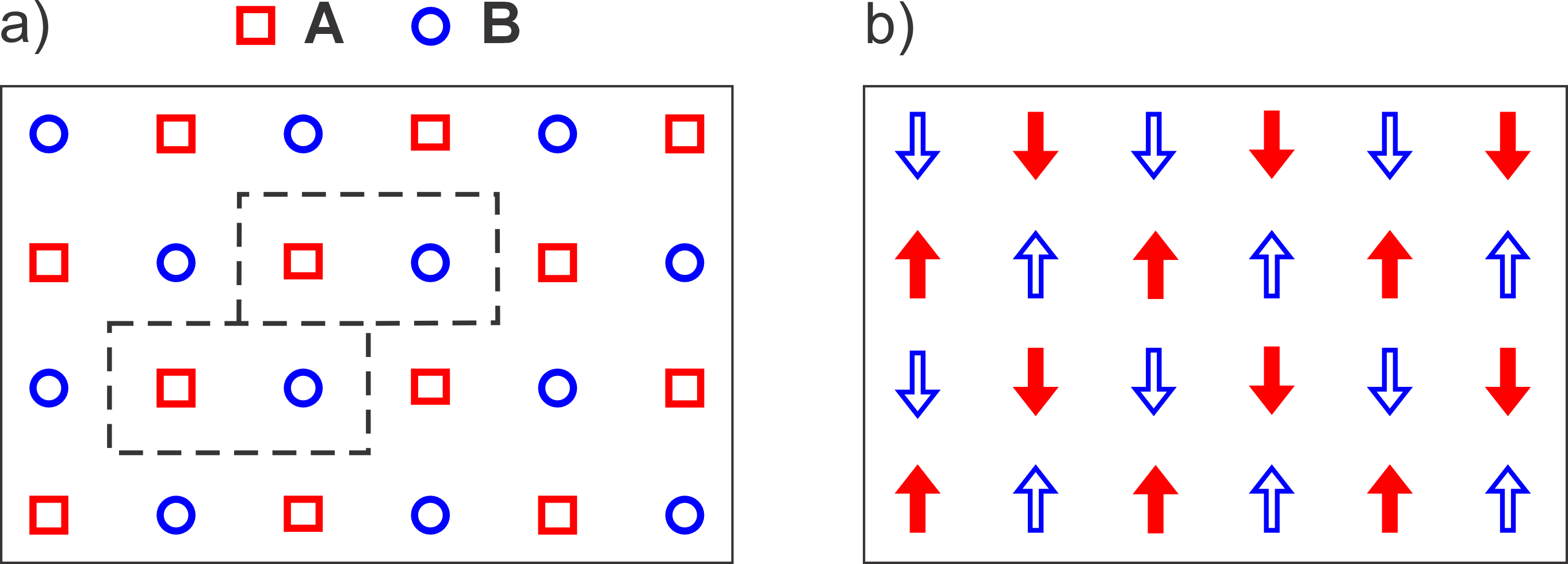}
\caption{(a) A and B sublattices in the $xy$ plane. The dashed lines indicate the two-site unit cells. (b) Stripe order composed of interpenetrating Ne\'el orders on the two sublattices. }
\label{fig:sublattices}
\end{figure}

To characterize the stripe order that occurs in the layered $J_1-J_2$ Ising model,
\begin{equation}
H_0 = - J_1 \sum_{\langle ij \rangle} S_i S_j  - J_2 \sum_{\langle\langle ij \rangle\rangle}  S_i S_j - J_\perp \sum_{\langle ij \rangle_\perp}  S_i S_j~,
\label{eq:H0_sup}
\end{equation}
for positive $J_1$ and negative $J_2$ with $|J_2|/J_1>1/2$, it is convenient to define two interpenetrating square sublattices A and B in the $xy$ plane, as shown in Fig.\ \ref{fig:sublattices}(a).
The sublattices repeat periodically in the $z$ direction. In the following, it is convenient to divide the lattice into two-site unit cells (each containing one A site and one B site), as indicated in the figure, and label the lattice sites not by their site index $i$ but by their unit cell index $\nu$ and the sublattice labels A or B.

In the low-temperature phase, the spins on each of the sublattices develop Ne\'el order, illustrated in Fig.\ \ref{fig:sublattices}(b). It is characterized by the order parameters
\begin{eqnarray}
\psi_A &=&  \frac 2 {N} \sum_{\nu} \psi_{A,\nu} = \frac 2 {N} \sum_{\nu}  (-1)^{y_\nu} S_{A,\nu}~,\\
\psi_B &=&  \frac 2 {N} \sum_{\nu} \psi_{B,\nu} = \frac 2 {N} \sum_{\nu}  (-1)^{y_\nu } S_{B,\nu}~,
\label{eq:pisapisb}
\end{eqnarray}
where $y_\nu$ is the $y$ coordinate of unit cell $\nu$, and $N/2$ is the number of unit cells in the system.
The state is degenerate w.r.t.\ the relative orientation of the spins on the two sublattices. Depending on this relative orientation, the spins form stripes in the $x$ direction, as in Fig.\ \ref{fig:sublattices}(b), or in the $y$ direction. The stripe-ordered low-temperature state can thus be completely characterized by a two-component primary order parameter $(\psi_A,\psi_B)$  which features $Z_2 \times Z_2$ symmetry. The stripe order parameters $\psi_x$ and $\psi_y$ defined in eq.\ (2) of the main text are linear combinations, $\psi_x = (\psi_A + \psi_B)/2$ and  $\psi_y = (\psi_A - \psi_B)/2$. Thus, the two-component primary order parameter characterizing the stripe phase can, equivalently, be written as $\psi= (\psi_x, \psi_y)$.

To characterize the vestigial order that breaks the symmetry between the $x$ and $y$ directions of the lattice (irrespective of the presence or absence of magnetic long-range order), we construct a composite order parameter that measures the relative orientation of the two sublattices. Following the framework of Ref.\ \cite{FernandesOrthSchmalian19S}, it can be defined as the product of the order parameter components  $\psi_{A,\nu}$ and $\psi_{B,\nu}$ in the same unit cell,
\begin{equation}
\bar\eta =\frac 2 {N} \sum_{\nu} \psi_{A,\nu} \psi_{B,\nu}  = \frac 2 {N} \sum_{\nu} S_{A,\nu} S_{B,\nu}
\label{eq:eta_bar}
\end{equation}
The nematic order parameter $\eta$ used in the main text,
\begin{equation}
\eta =\frac 1 N \sum_i \eta_i = \frac 1 {4N} \sum_i S_i( S_{i,+x} + S_{i,-x} - S_{i,+y} -  S_{i,-y}  )~,
\end{equation}
is a symmetrized version (in the $x$ and $y$ directions) of this composite order parameter.


\section{S2. Random-field disorder}

The quenched random fields  $\phi_i$ employed in the Monte Carlo simulations  are uniformly distributed on the interval $(-W,W)$. They are short-range correlated with a correlation length $\xi_d$, i.e., their covariance decays exponentially,
\begin{equation}
 C_\phi (\mathbf{r}_{ij}) = [ \phi_i \phi_j ]_\mathrm{rf} \propto \exp(-|\mathbf{r}_{ij}|/\xi_d).
\label{eq:corrfun}
\end{equation}

To generate the random-field values for each disorder configuration, we use
the Fourier filtering method \cite{MakseHavlinSchwartzStanley1996S}. First, uncorrelated Gaussian random numbers $\zeta_i$ of zero mean and unit variance are created on each site of the lattice. This uncorrelated random field is transformed to momentum space using a fast Fourier transformation. Its Fourier transform  is multiplied by the square root of the Fourier transform of the target correlation function,
$\tilde{\phi}_G(\mathbf{q}) = |\tilde C_\phi(\mathbf{q})|^{1/2} \tilde{\zeta}(\mathbf{q})$,
The inverse Fourier transformation of  $\tilde{\phi}_G(\mathbf{q})$  produces a Gaussian random field with the desired spatial correlations. The Gaussian field is finally mapped to a random field $\phi_i$ that is uniformly distributed on \([-W,W]\) using the complementary error function.

\begin{figure}[t]
  \includegraphics[width=\columnwidth]{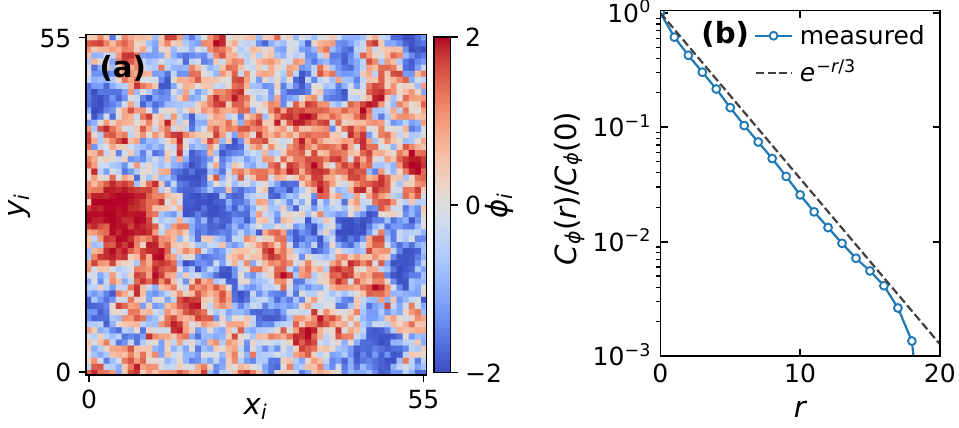}
  \caption{Verification of the correlated random-field construction for an $L=56$
  sample with $\xi_d=3$.  (a)  Two-dimensional slice of the generated field $\phi_i$ for a single disorder configuration.  (b) Measured (normalized) radial
  correlation function $C_\phi(r)$ compared to  the target exponential form  $\exp(-r/3)$. The data are averaged over two samples with $L=56$.   }
  \label{fig:phi-correlation-check}
\end{figure}

Figure \ref{fig:phi-correlation-check}(a)  shows an example of the resulting random field configuration. More precisely, it shows the random field values in a single $xy$ layer of a cubic sample with $L=56$ for $W=2$ and $\xi_d=3$.
The figure demonstrates the formation of small clusters of sites with similar $\phi_i$, consistent with short-range correlations. We have explicitly verified that the random field values have the desired covariance $C_\phi (\mathbf{r})$.   Figure \ref{fig:phi-correlation-check}(b) shows the (normalized) covariance  and compares it to eq.\ (\ref{eq:corrfun}). The measured correlations follow the target form well, but lie slightly below it. The small difference can be attributed to the mapping
of the random field values to a uniform distribution. The Fourier filtering method generates Gaussian random numbers with covariance  (\ref{eq:corrfun}), but the mapping slightly modifies the correlations.

\section{S3. Monte Carlo algorithm}

The interactions of $J_1-J_2$ Ising model (\ref{eq:H0_sup}) are frustrated in the parameter region of interest, $J_1>0$ and $J_2<0$. Therefore, cluster algorithms such as the Wolff algorithm \cite{Wolff89S} or the Swendsen-Wang algorithm \cite{SwendsenWang87S}  are not applicable to this system or, at least, not efficient. We therefore initially used the Metropolis algorithm \cite{MetropolisRosenbluthRosenbluthTellerTeller1953S}. The simulations were parallelized at the disorder configuration level, i.e., each disorder configuration was assigned to one MPI rank (CPU core).

This approach worked for smaller systems with sizes up to $L=40$ (even though the numerical effort to equilibrate a sample at $L=40$ was enormous).  For system sizes $L>40$, pure Metropolis simulations failed to equilibrate within the accessible simulation time.
Monte Carlo simulations of random-field systems are notoriously hard because the energy landscape contains many local minima
in which the simulation can get stuck. (Accordingly, for $L > 40$, we observed a strong dependence of the simulation results on the initial state).

To overcome these difficulties, we employed a parallel tempering Monte Carlo method
\cite{SwendsenWang1986S,Geyer1991S,HukushimaNemoto1996S}.  This algorithm simulates, for each disorder configuration, a ladder of $N_\mathrm{rep}$ replicas of the system at different temperatures. The temperature range is chosen such that equilibration is easy at the highest temperatures, whereas the temperatures of interest are at the lower end of range.  Local Metropolis sweeps are used to update each replica. After every five sweeps, neighboring replicas on the temperature ladder attempt exchanges with probability $P_{\mathrm{swap}}=\min\{1,\exp[(\beta_i-\beta_j)(E_i-E_j)]\}$,
where $E_i$ and $E_j$ are the total energies of the two neighboring replicas at inverse temperatures $\beta_i$ and $\beta_j$.
The swap probabilities fulfill detailed balance, ensuring that all replicas approach thermal equilibrium at their respective temperatures.
The high-temperature replicas can easily overcome barriers between different minima in the energy landscape. Via the exchanges, they help the low-temperature replicas escape such local minima
\cite{EarlDeem2005S}.

The first parallel-tempering implementation was parallelized at the disorder configuration level, i.e., each disorder configuration was assigned to one MPI rank (CPU core). This CPU core simulated all replicas in the temperature ladder for this configuration. This algorithm was sufficient for lattices up to $L=40$, but it turned out to be too slow for $L=56$. To make larger sizes accessible, we modified the algorithm such that it parallelizes at the replica level. In this mode, a group of CPU cores  is assigned to one configuration, with each core evolving a single replica in the temperature ladder.
The final (production) data set analyzed in the main text uses parallel-tempering data with parameters listed in Table~\ref{tab:simulation-parameters}.
The source code and raw data can be found in the public GitHub repository \cite{J1J2PublicDataS}.
\begin{table*}[t]
\caption{Simulation parameters for the final (production) data set:
\texttt{pt\_prod\_main\_plus\_l56\_1Meq\_500k\_susc} (see Ref.\ \cite{J1J2PublicDataS}).  All runs used
\(J_1=-J_2=J_\perp=1\), random field strength \(W=2\), correlation length \(\xi_d=3\). Each parallel tempering (PT) cycle  consists of 5 Metropolis sweeps for each replica, followed by swap attempts between replica pairs having adjacent temperatures.}
\label{tab:simulation-parameters}
\begin{ruledtabular}
\begin{tabular}{lccccc}
size(s) & starts & configurations per start & \(N_{\rm rep}\) & \(T\) range & equil./meas. PT cycles \\
\hline
\(20,28,40\) &
\(x\)-stripe, \(y\)-stripe, hot &
1000 each &
20 &
\(8.0\)--\(11.5\) &
\(10^5/10^5\) \\
\(56\) &
\(x\)-stripe &
256 &
64 &
\(8.0\)--\(11.5\) &
\(10^6/5\times10^5\) \\
\(56\) &
\(y\)-stripe &
64 &
64 &
\(8.0\)--\(11.5\) &
\(10^6/5\times10^5\) \\
\(56\) &
hot &
62 &
64 &
\(8.0\)--\(11.5\) &
\(10^6/5\times10^5\)
\end{tabular}
\end{ruledtabular}
\end{table*}

\begin{figure}
    \includegraphics[width=\columnwidth]{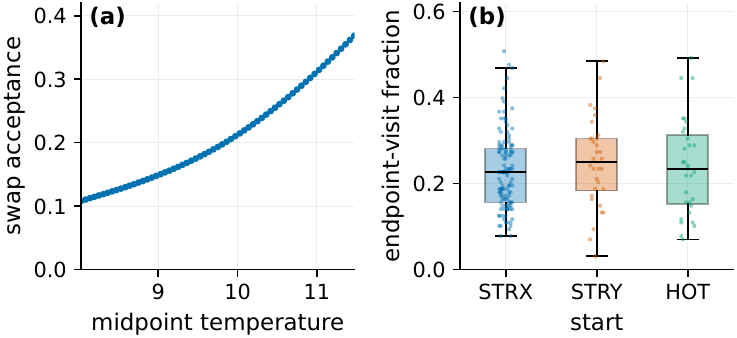}
    \caption{Parallel-tempering diagnostics for \(L=56\).
    (a) Mean replica swap acceptance between neighboring temperatures in the 64-replica ladder.
 (b) Fraction of replicas  that visited both the lowest and highest temperatures during the measurement phase, shown separately for $x$-stripe, $y$-stripe, and hot starts.
     Each point represents the average of two random-field configurations. The box plot indicates the mean, middle 50\% and total spread of the values over all disorder configurations in the production data set, see Table \ref{tab:simulation-parameters}.}
    \label{fig:pt-diagnostics-l56}
\end{figure}

Figure \ref{fig:pt-diagnostics-l56} illustrates the performance of the parallel-tempering algorithm for the largest system size, $L=56$. Panel (a) shows that average swap acceptance probability as a function of temperature.
The acceptance probability remains above 0.10 across the entire temperature range of interest and increases at higher temperatures. This implies that a given replica can move across the temperature ladder quite easily and overcome energy barriers. This is further illustrated in Fig.\  \ref{fig:pt-diagnostics-l56}(b) which shows that a significant fraction of replicas visit both ends of the temperature ladder.

\section{S4. Equilibration checks:}

To confirm that our simulations reach thermal equilibrium, we compare simulation results for three different starting states, as is often done in Monte Carlo simulations. Specifically, we compare simulations with $x$-stripe, $y$-stripe, and hot starts.
In case of an $x$-stripe start, the spins of the lattice are initialized such that perfect stripes form parallel to the $x$-axis, meaning \(\langle\psi_x\rangle =1 \) and \(\langle\psi_y\rangle =0 \).
A $y$-stripe start is analogous, with the stripes forming parallel to the $y$-axis, meaning \(\langle\psi_y\rangle=1\) and \(\langle\psi_x\rangle =0 \).
In the case of a hot start, the spins are initialized randomly, so both \(\langle\psi_x\rangle\) and \(\langle\psi_y\rangle\) are small initially.
If the system is equilibrated, the observables from all starting states should agree with each other within their error bars.

\begin{figure*}
  \includegraphics[width=\textwidth]{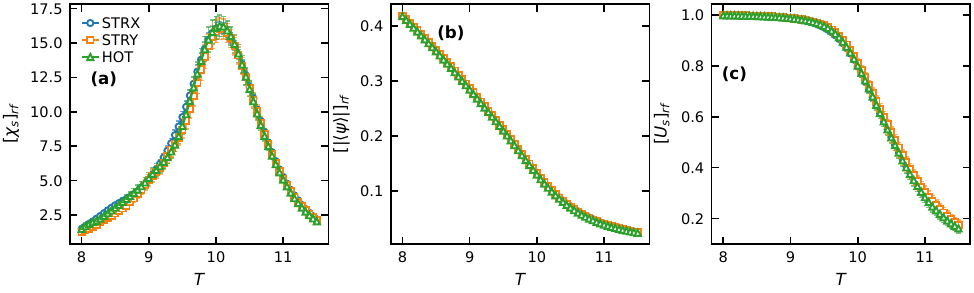}
  \caption{Equilibration check of three observables for \(L=56\).  The
  panels compare (a) the stripe susceptibility \(\chi_s\), (b) the stripe
  order parameter magnitude \(|\langle\psi\rangle|\), and (c) the stripe
  Binder cumulant \(U_s\) for simulations initialized from \(x\)-stripe,
  \(y\)-stripe, and hot starts. The data are averages over  all disorder configurations in the production data set, see Table \ref{tab:simulation-parameters}.}
  \label{fig:equilibration-observable-overlays-l56}
\end{figure*}

Figure \ref{fig:equilibration-observable-overlays-l56} presents the results of such an analysis.
It shows the temperature dependencies of the stripe susceptibility $\chi_s$, the stripe order parameter $|\psi|$, and the stripe Binder cumulant $U_s$ for runs with $x$-stripe, $y$-stripe and hot starts. The data for the three starts agree within the error bars, except
 at the very lowest temperatures for \(\chi_s\) where the data indicate a minor start-dependence.

\begin{figure*}[t!]
   \includegraphics[width=\textwidth]{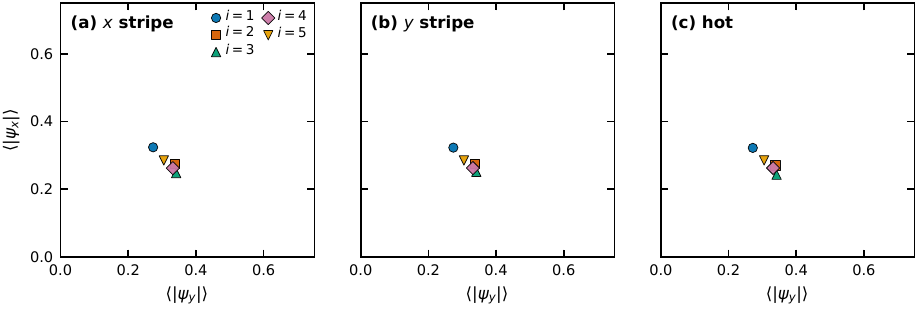}
  \caption{Equilibration check for individual disorder configurations at \(L=56\) and $T=8.0$: Stripe order parameter components  $\langle|\psi_x|\rangle$  and
  $\langle|\psi_y|\rangle$  for five individual random-field   configurations, distinguished by color and symbol.  (a)  \(x\)-stripe start, (b) \(y\)-stripe start, and (c) hot start.}
  \label{fig:equilibration-start-scatter-l56}
\end{figure*}

The results in Fig.\ \ref{fig:equilibration-observable-overlays-l56}  confirm the equilibration of disorder-averaged observables. In addition, we can also test the equilibration of individual disorder configurations. Figure \ref{fig:equilibration-start-scatter-l56} shows the stripe order parameter values  \(\langle|\psi_x|\rangle\) and \(\langle|\psi_y|\rangle\) for several individual disorder configurations as dots in the  \(\langle|\psi_x|\rangle - \langle|\psi_y|\rangle\) plane.
If the system is equilibrated, all three starts should lead to the same point in this plane for a given disorder configuration, but different disorder configurations should yield different points. The figure shows that this is indeed the case with high accuracy, providing a sensitive test of equilibration.

\begin{figure*}[t!]
  \includegraphics[width=\textwidth]{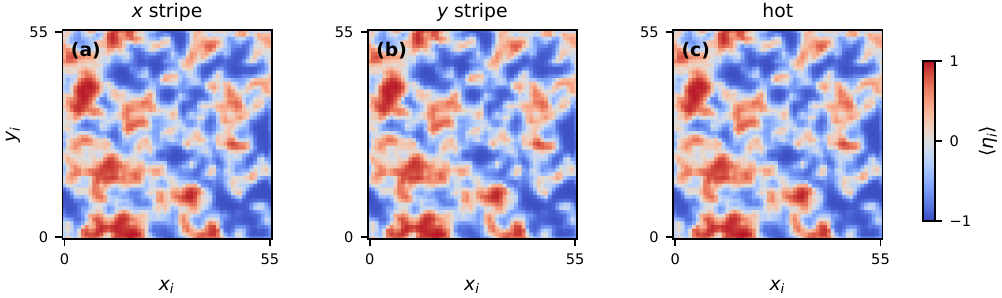}
  \caption{Equilibration check for local nematic order parameter at \(L=56\) and $T=8.0$:  The panels show
  the local nematic order parameter \(\langle\eta_i\rangle\)   in the same \(z=32\) layer of the same random-field configuration after
  (a) \(x\)-stripe start, (b) \(y\)-stripe start, and (c) hot start.}
  \label{fig:equilibration-nematic-heatmaps-l56}
\end{figure*}

A further analysis can be performed by comparing the local nematic order parameters $\langle \eta_i \rangle$ resulting from runs with different starts.
Figure \ref{fig:equilibration-nematic-heatmaps-l56} shows $\langle \eta_i \rangle$ in a single layer of one disorder realization
for three different starts.
The domain patterns are virtually indistinguishable between the three runs, confirming that the local nematic texture has equilibrated.

\end{document}